\documentclass[11pt,preprint]{aastex}
\usepackage{amsmath} 
\usepackage{amssymb} 
\usepackage{verbatim}
\usepackage{graphicx}
\usepackage[percent]{overpic}
\usepackage{subfigure}
\usepackage{esint}
\usepackage{url}
\usepackage{natbib}
\usepackage{mycommands}
\begin{document}

\title{PICsar2D: Public Release}

\begin{abstract}
{\it PICsar2D} is a 2.5D relativistic, electromagnetic, particle in cell code designed for studying the pulsar magnetosphere. The source code and a suite of {\it Python} analysis routines can be downloaded from ``https://github.com/astromb/PICsar2D.git". Additionally, the repository includes a step-by-step tutorial for compiling the code, running it, and analyzing the output.  This article is devoted to several new algorithmic advances and numerical experiments. These include a new pair injection prescription at the pulsar surface, a comparison of different pair injection techniques, a discussion of particle trajectories near the pulsar Y-line, and performance optimization of the code.
\end{abstract}

\author{Mikhail A. Belyaev}
\affil{Astronomy Department, University of California
    Berkeley \\ mbelyaev@berkeley.edu}

\section{Introduction}

{\it PICsar2D} is a particle in cell (PIC) code that was developed for studying the pulsar magnetosphere \citep{BelyaevPIC}. The code solves Maxwell's equations on a mesh with a current source term provided by particles. Electric and magnetic fields are advanced in time using standard finite difference time domain techniques \citep{Yee}, and currents are deposited to the grid via either the Villasenor-Buneman \citep{VillasenorBuneman} or Esirkepov \citep{Esirkepov} algorithms. The particle positions and velocities are leapfrogged in time using either the relativistic version of the Boris mover \citep{Boris} or the Vay mover \citep{Vay}. The advantage of the Vay mover is that it accurately captures the ExB particle drift, even when the particle Larmor frequency is not resolved in time. The code is charge-conservative (to machine precision), and therefore no divergence cleaning is required. A current filtering algorithm that preserves the charge-conservative property of the code is used to suppress high-frequency noise. Additional features of the code include radiation absorbing boundary conditions and variable grid spacing in the radial and theta directions.

{\it PICsar2D} has already been used to simulate the pulsar magnetosphere \citep{BelyaevPIC2}. Here, we focus on new results pertaining to numerical experiments with pair production and particle motion near the Y-line. Additionally, we present performance optimization techniques that have been used in tuning the public release of the code. The source code and analysis routines used to generate the plots in this paper and in \cite{BelyaevPIC,BelyaevPIC2} can be downloaded from ``https://github.com/astromb/PICsar2D.git". The repository also includes a step-by-step tutorial that details how to compile and run the code, as well as use the analysis routines.  

\section{Numerical Experiments with New Pair Production Algorithm}

\subsection{New Pair Production Algorithm}

There have already been a number of works investigating the pulsar magnetosphere with PIC \citep{SpitkovskyPIC,ChenBeloborodov,Cerutti,BelyaevPIC2,Philippov3D,PhilippovGR}. The numerical methods used in these studies are broadly consistent, but one major source of variation between them is the particle injection method. 

Charged particles are expected to be present in the pulsar magnetosphere for two reasons. First, particles are pulled off the neutron star, because they are unbound by the intense electromagnetic fields at the surface \citep{GoldreichJulian}. Second pair production of positrons and electrons (e.g.\ via gamma rays on magnetic field lines \citep{Sturrock}) creates a pair plasma in the vicinity of the pulsar surface. 

The particle injection prescription in \cite{BelyaevPIC2} consisted of two parts. The first part modeled the emission of charged particles from the neutron star via a surface charge emission algorithm. This algorithm uses Gauss's law to compute the amount of charge accumulated on the pulsar surface each timestep and releases a fraction of it into the magnetosphere. The second part modeled the presence of pairs in the magnetosphere. To mock up a pair plasma, \cite{BelyaevPIC2} injected pairs throughout a spherical volume of the simulation domain with radius of order the light cylinder radius in such a way that $\bfE \cdot \bfB$ was forced towards zero everywhere throughout this volume. This volumetric injection prescription was ad hoc, but it did achieve its intended purpose of generating a pair plasma around the pulsar. 

\cite{Cerutti} implemented a different pair injection algorithm. They injected pairs at the surface of the pulsar and gave them a small initial outward velocity\footnote{The small initial velocity can be physically justified by the fact that pair production by gamma rays on magnetic field lines leads to initial pair velocities that are outwardly directed along field lines.}. This injection method fills the magnetosphere with pairs that come exclusively from the surface of the neutron star. Thus, it can be viewed as a subgrid model for a surface pair cascade.

The algorithm of \cite{Cerutti} produces pairs everywhere on the surface of the neutron star. However, \cite{Beloborodov,TimokhinArons} have shown that a surface pair cascade is present only if certain criteria are satisfied locally: for surface pair production to occur, {\it either} the four-current is spacelike, {\it or} it is timelike but the current density is in the sense of the Goldreich-Julian density flowing into the star. If the four-current is timelike {\it and} the current density is in the same sense as the Goldreich-Julian density flowing away from the star, pairs are not produced locally. The Goldreich-Julian density (in flat spacetime) can be written as
\ba
\rho_\text{GJ} = - \frac{\bfOmega \cdot \bfB}{2 \pi c} \ \ \ \text{(Gaussian units)}.
\label{rhoGJ}
\ea

\cite{BelyaevParfrey,GrallaPhilippov} have shown that general relativity (GR), and in particular frame dragging, are important for determining the sign of the four-current at the polar cap. Since the latter is used in the pair production criteria, GR plays an important role with regards to surface pair production at the polar cap. 
However, frame dragging and GR do not qualitatively affect most of our results, many of which carry over even if we assume spacetime is flat. Thus, we shall try to point out explicitly where GR and frame dragging do make a difference when discussing the results of numerical experiments.

The new pair injection algorithm is similar to the surface pair injection scheme of \cite{Cerutti},  except that pairs are only injected at the pulsar surface if the pair production criteria of \cite{Beloborodov,TimokhinArons} are locally satisfied. In order to control the pair multiplicity, we inject up to a specified multiplicity within one cell of the surface of the neutron star each timestep. The utility of the new pair injection algorithm is that it provides a physically-motivated subgrid implementation of surface pair injection. Thus, it provides an alternative to direct numerical simulation of the pair cascade \citep{PhilippovGR}. The latter approach is computationally expensive in a global simulation because of the extreme ratio between the time and length scales of pair production compared to the time and length scales of magnetospheric physics.

\subsection{Numerical Experiments with Surface Pair Production}

We now perform numerical experiments that run the gamut of different options for surface pair injection within {\it PICsar2D}. These experiments not only demonstrate different ``states" that are possible in the pulsar problem with PIC, but also highlight the effect that different pair production prescriptions have on the solution. These numerical experiments can be run using the publicly available version of the {\it PICsar2D} code and require only modest computing resources.

We perform four simulations of the aligned rotator with different pair production prescriptions. Features common to each of the simulations are 1) surface charge emission is turned on for the entire simulation; 2) the grid dimensions are $N_r \times N_\theta = 513 \times 257$ ($N_r$ and $N_\theta$ give the number of grid edges which is one more than the number of grid cells); 3) the grid is logarithmic in the radial direction and has extent $ 1 < r/r_* < 54.5$; 4) an equal area theta grid is used in the $\theta$ direction to reduce particle shot noise near the poles \citep{BelyaevPIC} (the simulation extent in $\theta$ is to the poles, $0 < \theta < \pi$); 5) the light cylinder radius is at $R_\text{lc} = 4 r_*$; 6) the simulations contain only electrons and positrons (i.e.\ no ions or other particle species are included); 7) the simulations are for an aligned rotator; 8) the initial magnetic field structure is a dipole; 9) the Vay particle pusher is used rather than the Boris pusher; 10) Esirkepov current deposition is used rather than Villasenor-Buneman (although this essentially has no effect on the results); 11) the current is filtered twice using the same filtering algorithm as in \citep{BelyaevPIC} to reduce high frequency noise.

Our four numerical experiments are labeled A1, A2, D1, and D2, respectively, and they differ only in the pair production prescription. All simulations are run for a duration of $t = 3.730 P_*$, where $P_* = 2\pi/\Omega_*$ is the pulsar period with the exception of simulation D2, which is run for four times longer. Simulation D1 has no pair production, only surface charge emission, similar to \cite{KrausePolstorffMichel,SpitkovskyPIC}. Simulation A1 has pair production on all field lines for all time, similar to \cite{Cerutti}. Simulation D2 has pair production on all field lines turned on for $t < 3.730 P_*$, but pair production is turned off entirely for $t > 3.730 P_*$, leaving only surface charge emission. Simulation A2 has pair production on all field lines turned on for $t < 1.865 P_*$ but for $t > 1.865 P_*$ we use our new injection algorithm that only injects pairs on field lines satisfying the criteria of \cite{Beloborodov, TimokhinArons}. We take the frame dragging rate in simulation A2 to be $\omega_\text{LT}(r_*) = .21 \Omega_*$ as in \cite{BelyaevParfrey} when computing the four current over the polar cap, which enters into the pair production criteria. 

\begin{figure}[!t]
\centering
\includegraphics[width=.49\textwidth]{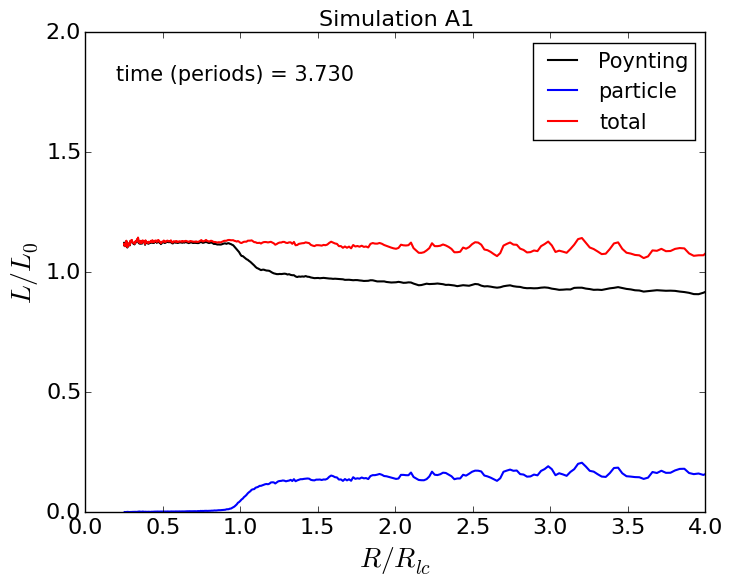}
\includegraphics[width=.49\textwidth]{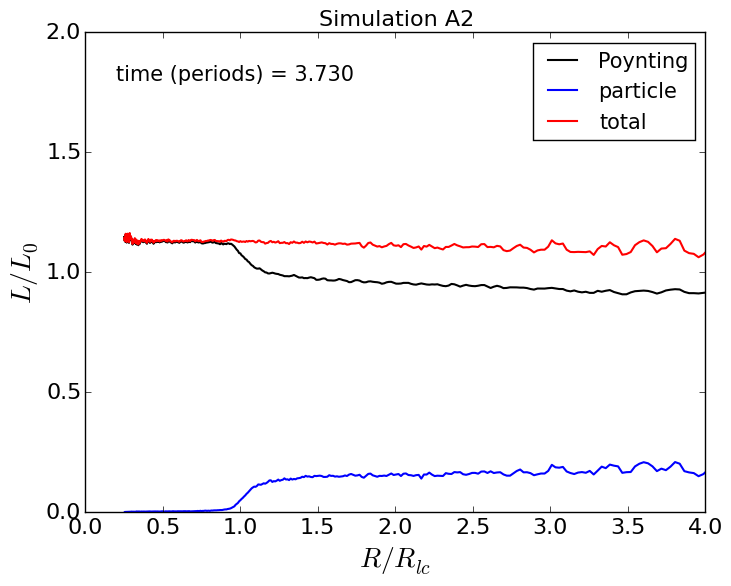}
\includegraphics[width=.49\textwidth]{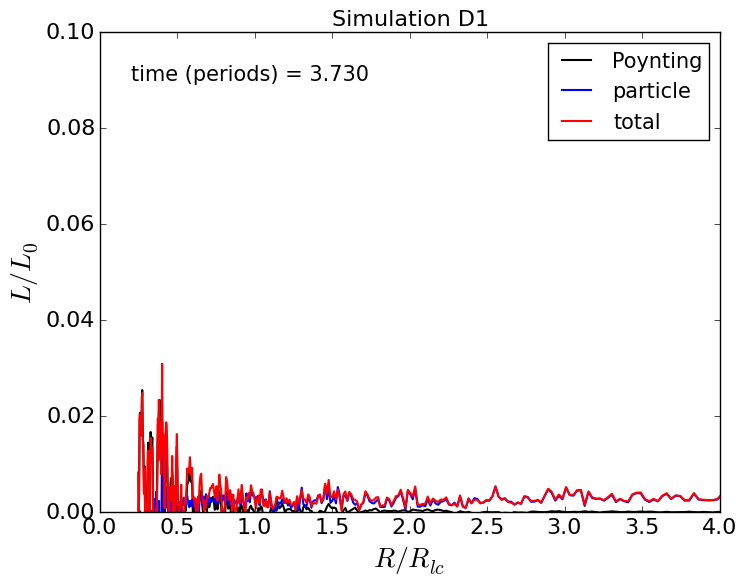}
\includegraphics[width=.49\textwidth]{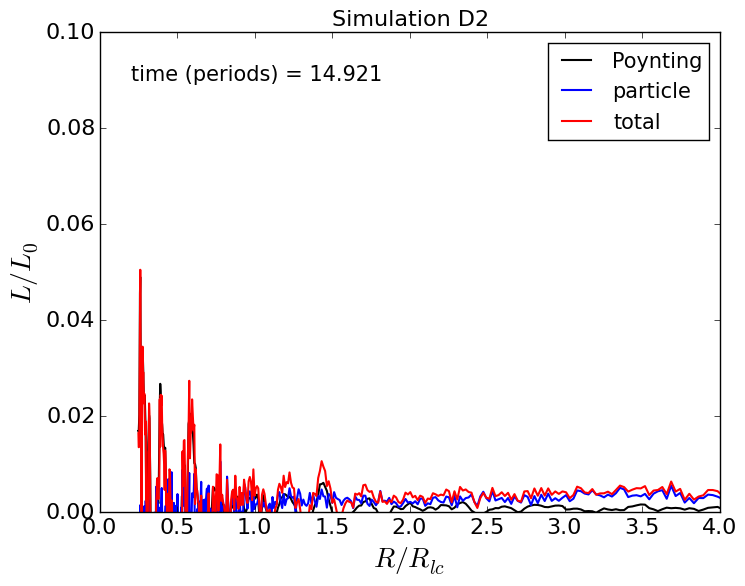}
\caption{Aligned rotator spindown luminosity for simulations A1, A2, D1, \& D2. The red line shows the total spindown luminosity, and the black and blue lines show the respective contributions from the electromagnetic and particle components.}
\label{poynt_plot}
\end{figure}

Note that in our simulations that include pairs, we only model surface pair production. We do not provide models for any pair production processes that have been suggested for the outer magnetosphere. Thus, any particles that are injected into the simulation domain are injected within one grid cell of the pulsar surface. Because there are no particles within the simulation domain initially, this means all particles in all simulations originate very near the surface of the neutron star. 

The different prefix labels, ``A" or ``D" for each simulation, denote whether the pulsar is in an active or a dead state at the end of the simulation. An active state means that the spindown luminosity is comparable to the spindown luminosity of the force-free aligned rotator \citep{CKF,SpitkovskyFF}, and a dead state means that the spindown luminosity is much lower than this. Fig.\ \ref{poynt_plot} shows the total spindown luminosity as a function of radius, as well as the individual contributions from the electromagnetic component (Poynting flux integrated over a spherical shell) and the particle component. The spindown luminosity is normalized to that of the force-free aligned rotator ($L_0$) and is measured at the end of each simulation. 

Active simulations have a spindown luminosity comparable to the force-free case, whereas dead simulations have a spindown luminosity that is effectively zero (note the different scale on the $y$-axis between the active and dead cases in Fig.\ \ref{poynt_plot}). However, the spindown luminosity is not exactly equal to the force-free luminosity, even for active simulations. Moreover, there is conversion of electromagnetic energy to particle kinetic energy beyond the light cylinder in the active simulations, which is not possible in the force-free limit. Both of these effects arise with PIC because the force-free approximation breaks down over a finite volume of the simulation domain. 

\begin{figure}[!t]
\centering
\includegraphics[width=.49\textwidth]{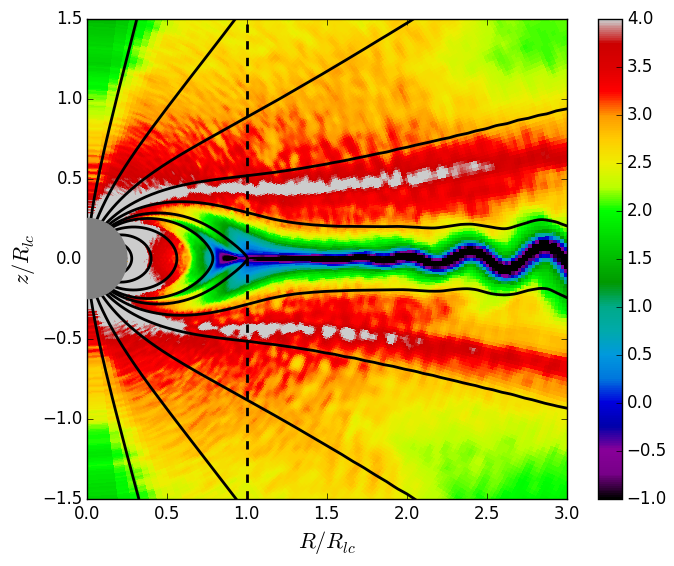}
\includegraphics[width=.49\textwidth]{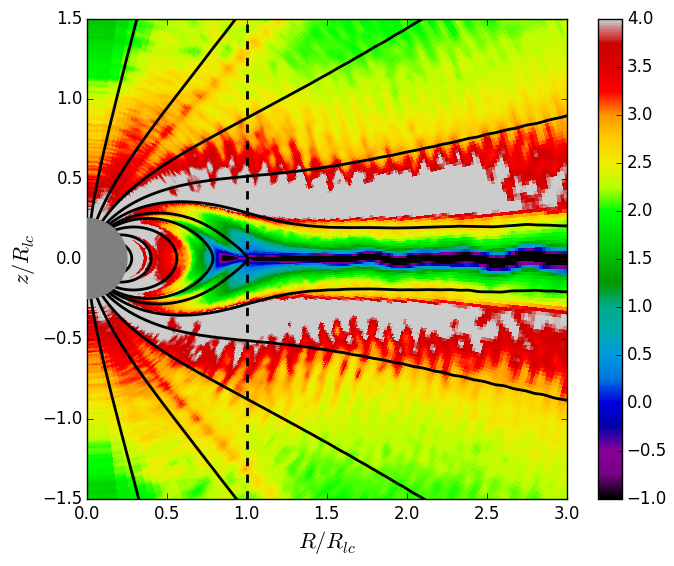}
\caption{Logarithm of the magnetization parameter, $\sigma$, for simulations A1 (left panel) and A2 (right panel) at the end of the simulation. The black curves are the magnetic field lines, and the gray inner circle is the neutron star. Note that just because a region has a high magnetization does not mean it is force-free, since magnetic field in vacuum has infinite magnetization. In fact, the regions on either side of the current layer in simulation A2 are vacuum gaps.}
\label{mag_plot}
\end{figure}

The magnetization is a useful parameter that probes the ratio between the magnetic and particle energy densities and determines how close to force-free the simulation is throughout the domain. Fig.\ \ref{mag_plot} shows the magnetization for simulations A1 and A2, where we define
\begin{align}
\label{magnetization}
\sigma \equiv \frac{B^2}{4 \pi (\gamma_+ n_+ + \gamma_- n_-) m c^2} \ \ \ \text{(Gaussian units)},
\end{align}
and $\gamma_+$, $n_+$ and $\gamma_-$, $n_-$ are the average gamma factors and densities of the positrons and electrons, respectively.  Near the Y-line and in the equatorial current layer, the magnetization is low implying that particle energy dominates magnetic energy. Particle inertia is important in these regions and tends to open up magnetic field lines. Because the spindown luminosity depends sensitively on the exact location of the Y-line as $L \sim L_0 (R_\text{Y}/R_\text{LC})^{-4}$, moving the Y-line radially inward by only 5\% amounts to a 20\% increase in the pulsar spindown luminosity. 

Next, we consider the reason why simulations D1 and D2 lead to dead pulsar states. Because the magnetospheric plasma is generally well-magnetized, currents flow along magnetic field lines in the meridional plane. In the active pulsar case, a return current flows at the boundary between the zone of open and closed field lines inside the light cylinder. The return current layer requires both electrons and positrons to be present for two reasons. First, the four-current in the return current layer is spacelike, and hence requires charge carriers of both signs moving in opposite directions. Second, field lines that thread the return current layer ``turn over" in the meridional plane ($B_z$ changes sign at some point on the field line), which implies that the Goldreich-Julian density (equation [\ref{rhoGJ}]) also changes sign on these field lines. Emission of surface charge alone provides particles of a single sign, {\it either} positive or negative, which depends on the sign of $\rho_\text{GJ}$ at the footpoint of the magnetic field line. Thus, for field lines that turn over, surface charge emission cannot supply plasma past the turn over point. For these reasons, pulsars with only surface charge emission (no pair production) cannot support a return current layer and are dead.

\begin{figure}[!t]
\centering
\includegraphics[width=.37\textwidth]{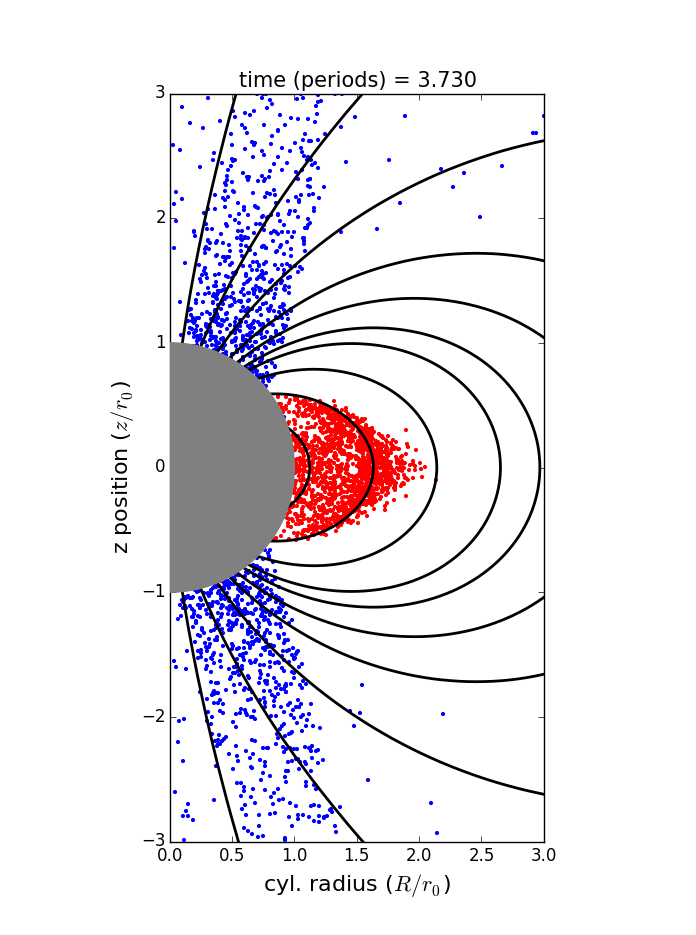}
\includegraphics[width=.59\textwidth]{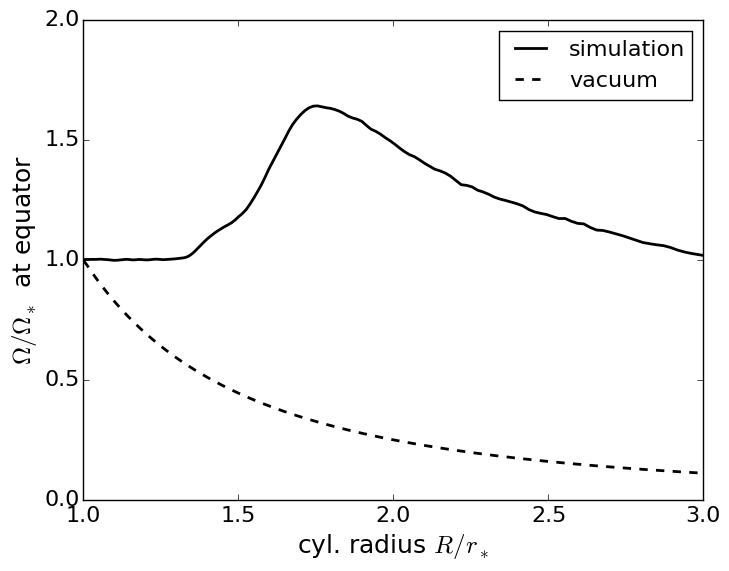}
\caption{Left panel: Downsampled snapshot of the spatial distribution of particles (positrons are red, electrons are blue) for simulation D1 at the end of the simulation. The black curves are magnetic field lines, and the gray sphere is the neutron star. Note that each point representing a particle is a toroidal ring in three dimensions due to axisymmetry. The white regions that have no particles are vacuum gaps. Right panel: Rotation profile in the equatorial plane normalized by the angular velocity of the neutron star, $\Omega_*$. The solid black curve is calculated at the end of the simulation and shows the rotation profile corresponding to the distribution of charge shown in the left panel. The dashed curve is the initial rotation profile which corresponds to vacuum outside the neutron star (no particles).}
\label{dead_fig}
\end{figure}

Because dead pulsars do not support return currents, the region of space around a dead pulsar is typically referred to as an ``electrosphere" rather than as a magnetosphere. Electrospheres were studied extensively by \cite{KrausePolstorffMichel,MichelLi}, who found a space charge separated plasma characterized by domes of negative charge near the poles and a torus of positive charge near the equatorial midplane\footnote{The only difference between the aligned and anti-aligned rotator as regards the electrosphere is that the sign of charge carriers is flipped in the domes and in the torus.}. The left panel of Fig.\ \ref{dead_fig} shows the charge distribution in the electrosphere at the end of simulation D1. Note that despite the large vacuum gaps in the magnetosphere, $\bfE \cdot \bfB = 0$ over the surface of the neutron star. This is because the surface charge injection algorithm does not allow charge to build up on the pulsar surface.

The rotation profile in the equatorial plane (right panel of Fig.\ \ref{dead_fig}) can be used to check the simulation result. Particles on magnetic field lines that are immersed in plasma corotate with the neutron star by Ferraro's isorotation law (or equivalently by the frozen-in flux theorem). In the right panel of Fig.\ \ref{dead_fig}, the plasma in the equatorial plane corotates up to a point $R \lesssim 1.5 R_*$, and then departs from corotation for $R \gtrsim 1.5 R_*$.  The radius $R \approx 1.5 R_*$ corresponds to the crossing radius in the equatorial plane of the last field line that is completely immersed in plasma (left panel of Fig.\ \ref{dead_fig}). Magnetic field lines that cross the equatorial plane at a larger radius pass through a region of vacuum between the domes and the torus. Hence, the plasma on these field lines is no longer bound to corotate with the pulsar by Ferraro's isorotation law.

\begin{figure}[!t]
\centering
\includegraphics[width=.32\textwidth]{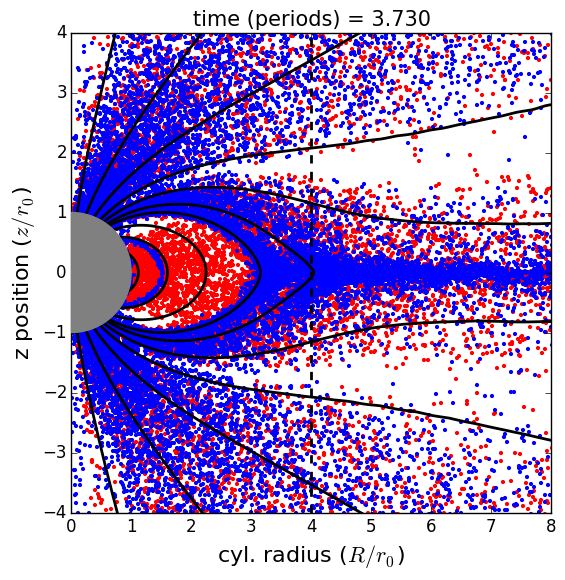}
\includegraphics[width=.32\textwidth]{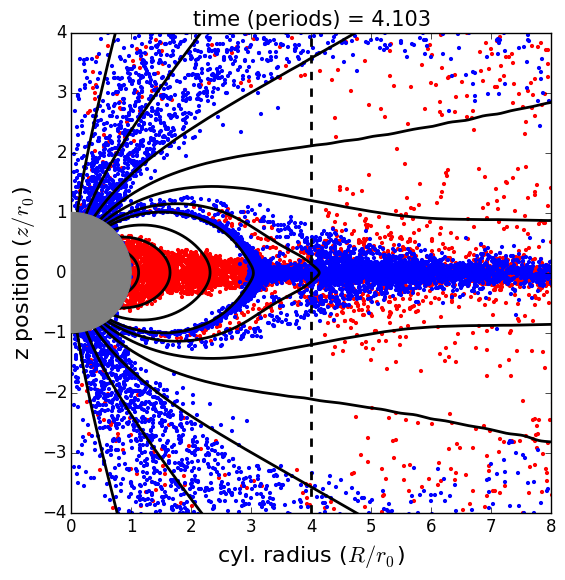}
\includegraphics[width=.32\textwidth]{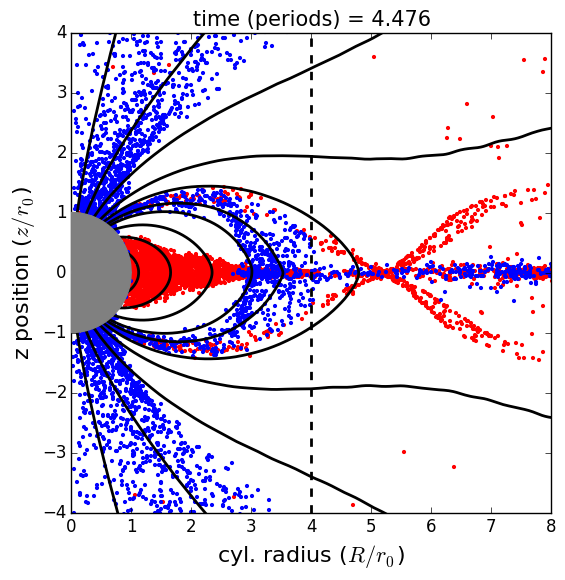}
\includegraphics[width=.32\textwidth]{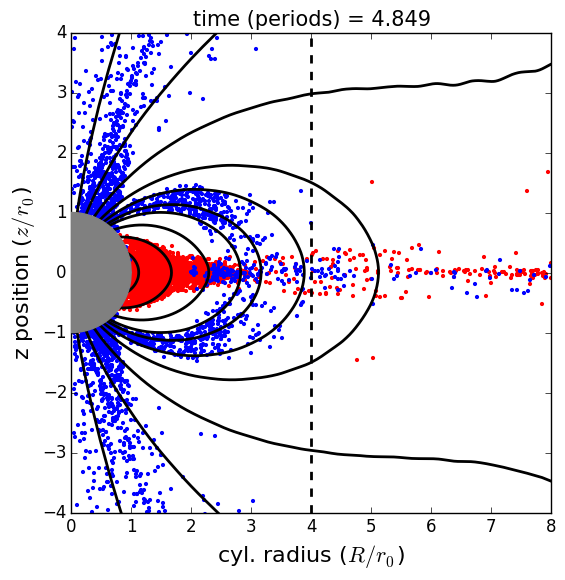}
\includegraphics[width=.32\textwidth]{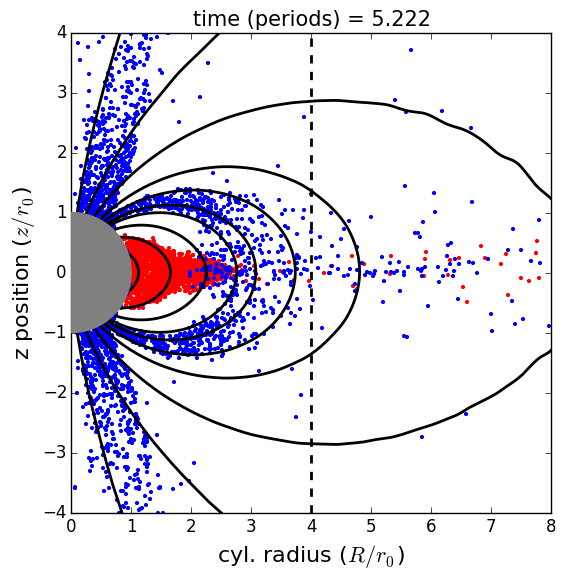}
\includegraphics[width=.32\textwidth]{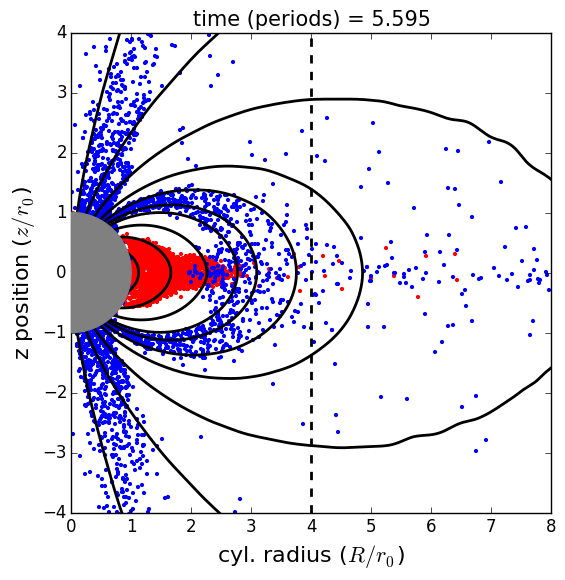}
\caption{Downsampled snapshots of the spatial distribution of particles (positrons red, electrons blue) for simulation D2. The black curves are magnetic field lines and the gray sphere is the neutron star. The upper left panel corresponds to the time right before pair production is switched off in the simulation. After pair production is turned off, an electrosphere develops on a timescale comparable to the pulsar rotation period. Notice how the open field lines close and contract to a dipole structure after the return current ceases to flow.}
\label{transit_fig}
\end{figure}

Simulation D1 is in a dead state for all time, but even an active pulsar transitions to a dead state if pair production is turned off entirely. Fig.\ \ref{transit_fig} shows the transition from an active to a dead state in simulation D2. The red points represent positrons and the blue points electrons. At $t= 3.730 P_*$ the magnetosphere is in an active state with plasma present everywhere, since pair production has been turned on since the star of the simulation. After pair production is turned off, current ceases to flow in the return current layer, and the pulsar circuit is quenched. The magnetosphere collapses to an electrosphere and the pulsar transitions to a dead state on a timescale comparable to the pulsar period. The spindown luminosity in simulation D2 is slightly elevated compared to simulation D1 for several rotational periods but settles down to the same low levels as in simulation D1 (compare the two bottom panels of Fig. \ref{poynt_plot}).

Given that turning off pair production entirely in simulation D2 rapidly leads to a dead state, it is interesting that simulation A2 remains in an active state after $t > 1.875P*$, when pair production is shut off on field lines not satisfying the pair production criteria of \cite{Beloborodov,TimokhinArons}. Perhaps not surprisingly, the reason why simulation A2 remains active has to do with the return current layer. Because the four-current in the return current layer is spacelike, the footpoints of the magnetic field lines along which the return current flows sustain pair production according to the criteria of \cite{Beloborodov,TimokhinArons}. Thus, the return current is not quenched in simulation A2 after the pair production criteria are imposed, and the simulation remains in an active state indefinitely.

\begin{figure}[!t]
\centering
\subfigure{\begin{overpic}
	[width=.48\textwidth]{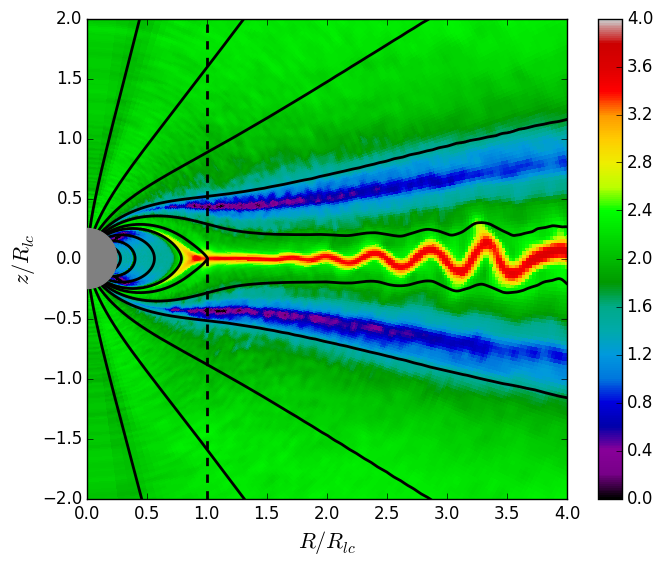}
	\put(15,76){\large particles per cell}
	\put(75,76){\large A1}
\end{overpic}}
\subfigure{\begin{overpic}
	[width=.48\textwidth]{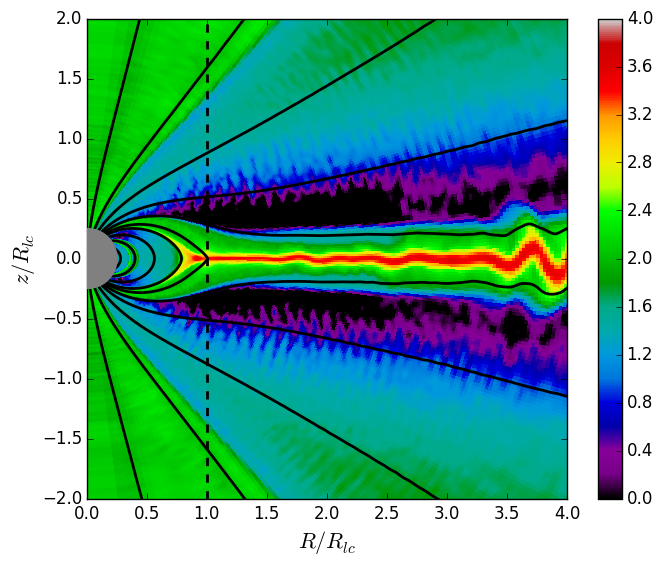}
	\put(15,76){\large particles per cell}
	\put(75,76){\large A2}
\end{overpic}}
\subfigure{\begin{overpic}
	[width=.48\textwidth]{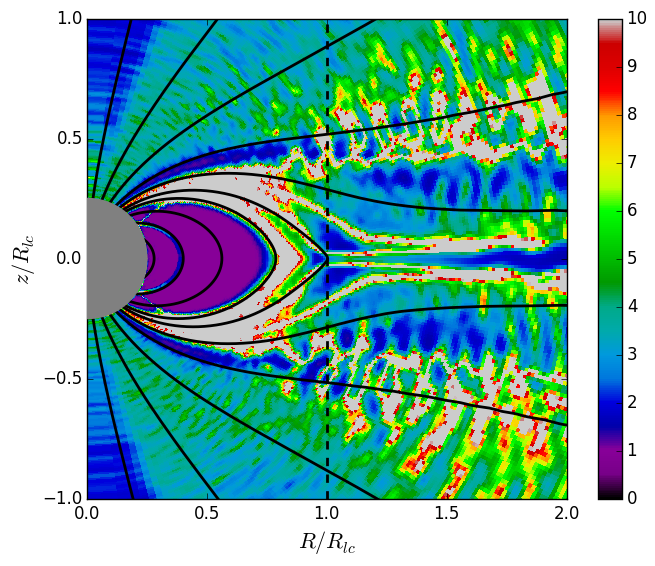}
	\put(15,76){\large pair multiplicity}
	\put(75,76){\large A1}
\end{overpic}}
\subfigure{\begin{overpic}
	[width=.48\textwidth]{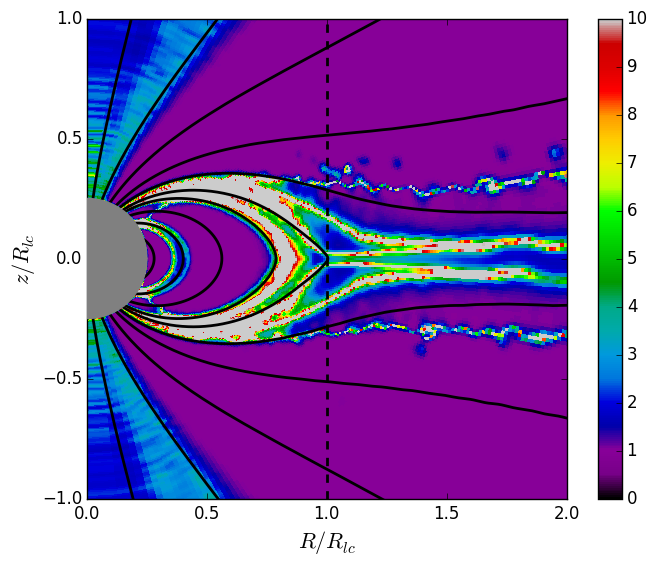}
	\put(15,76){\large pair multiplicity}
	\put(75,76){\large A2}
	\end{overpic}}
\caption{Upper row: Logarithm of the particles per cell at the end of the simulation ($t = 3.730P_*$) for simulations A1 (left) and A2 (right). The black wedge on either side of the current layer in simulation A2 is a vacuum gap that is devoid of both electrons and positrons. Lower row: Pair multiplicity for simulations A1 (left) and A2 (right). The large purple wedge for simulation A2 polewards of the the return current layer corresponds to field lines which do not support pair production and are populated by only a single sign of charge (electrons). The black curves are magnetic field lines, and the gray circle is the neutron star.}
\label{pairfig}
\end{figure}
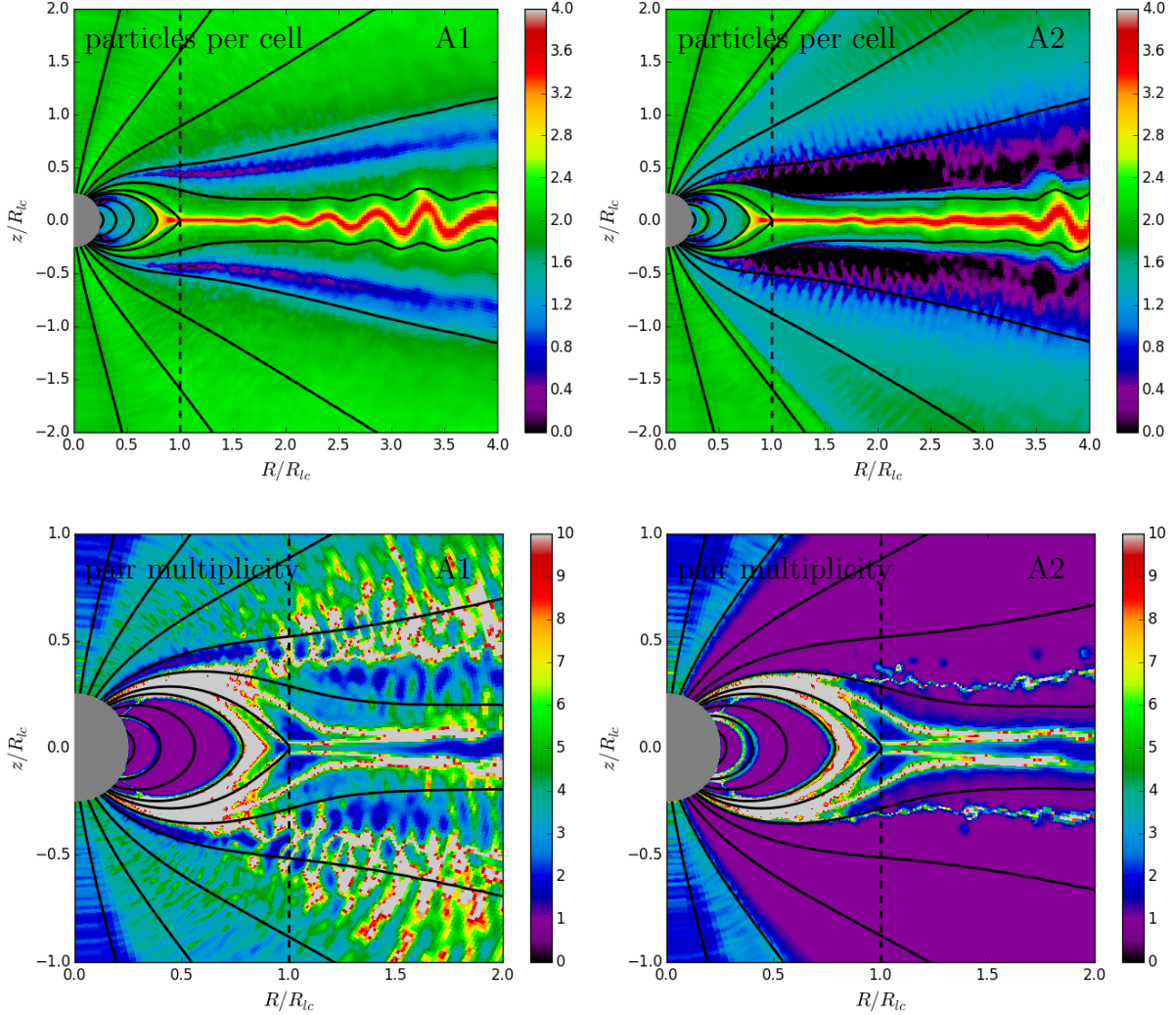

Even though simulations A1 and A2 both support active pulsars, there are important differences between them. For instance, there is no pair production within a large wedge in $\theta$ starting from the return current layer and extending towards the pole in simulation A2. As a result, vacuum gaps which are devoid of plasma are present on either side of the current layer. The upper panels of Fig.\ \ref{pairfig} show the particles per cell (electrons and positrons combined) for simulations A1 and A2. Vacuum gaps in simulation A2 are the black regions adjacent to the current layer. Field lines threading the vacuum gaps in simulation A2 are ones that do not support pair production according to the criteria of \cite{Beloborodov,TimokhinArons} {\it and also} turn over in the meridional plane. Surface charge emission only supplies a single sign of charge from the neutron star surface. Thus, these field lines contain no plasma\footnote{Although some particles may be present in the gaps, there is not enough plasma to short out the accelerating component of the electric field along magnetic field lines.} beyond the point where they turn over and the Goldreich-Julian density changes sign. 

At mid-latitudes, the field lines in simulation A2 do not support pair production, but they no longer turn over, so the Goldreich-Julian density has the same sign everywhere along the field line. In this case, surface charge emission can provide plasma everywhere on the field line, meaning these mid-latitude open field lines are populated with a single species electron plasma. Higher latitude field lines close to the polar axis support pair production, because the four-current (taking GR into account) is spacelike \citep{BelyaevParfrey,GrallaPhilippov}. 

The distinction between field lines that support pair production and those that do not becomes clear if we examine the pair multiplicity,
\ba 
\kappa \equiv \text{abs} \left( \frac{n_+ + n_-}{n_+ - n_-} \right) \ \ \ \text{(pair multiplicity)},
\label{pair_mult_def}
\ea
where $n_+$ and $n_-$ are the positron and electron densities, respectively. The lower panels of Fig.\ \ref{pairfig} show the pair multiplicity for simulations A1 and A2. In simulation A2, there is a large purple wedge at mid-latitudes that corresponds to field lines that do not support pair production and have pair multiplicity equal to unity. Near the polar axis and in the return current layer the pair multiplicity is higher, because surface pair production is turned on in these regions according to the criteria of \cite{Beloborodov,TimokhinArons}.

The frame dragging rate, $\omega_{\rm LT}(r_*)$, affects the $\theta$-extent of the high-latitude pair producing regions adjacent to the polar axis. As the frame dragging rate is decreased, the pair producing regions near the polar axis shrink in size, and for $\omega_{\rm LT} = 0$ (flat spacetime) they vanish altogether. However, even without pair production near the polar axis, there is still pair production in the return current layer. Thus, the pulsar remains active even in flat spacetime, because the presence of pairs in the return current layer is the fundamental ingredient that is necessary for the pulsar circuit to function.

\begin{figure}[!t]
\centering
\includegraphics[width=.48\textwidth]{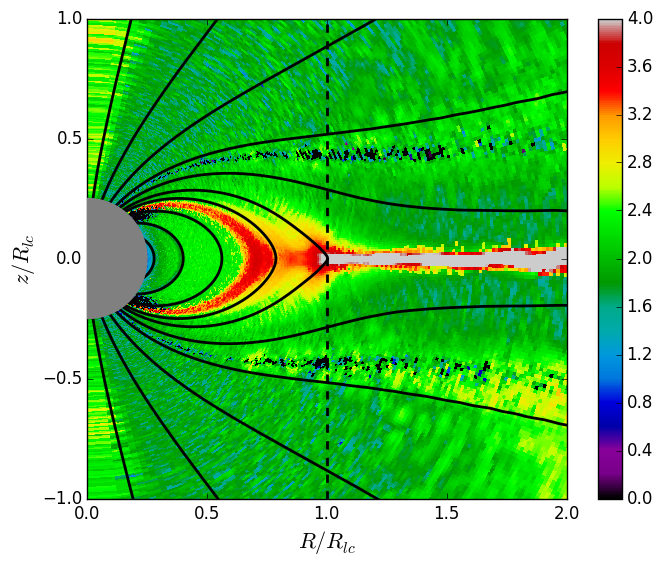}
\includegraphics[width=.48\textwidth]{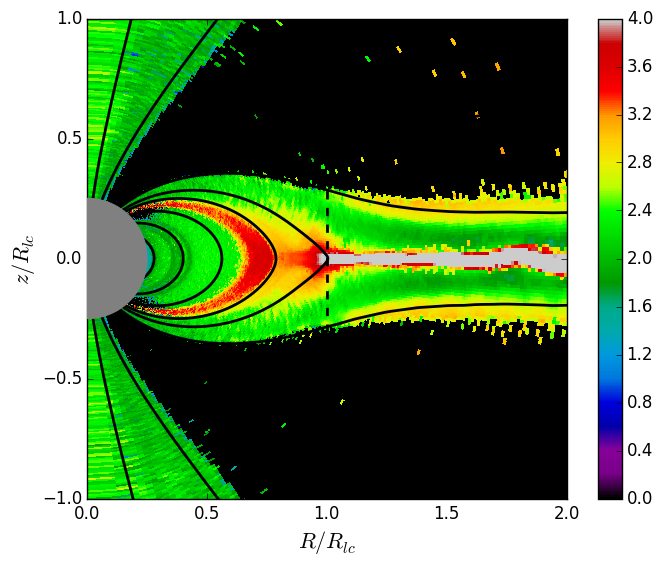}
\caption{Logarithm of the cell-averaged positron gamma factor at end of simulation A1 (left) and A2 (right). Particle acceleration occurs near the Y-line and in the current sheet. The large black wedge in the right panel is a region devoid of positrons (see discussion in text). The black curves are magnetic field lines, and the gray circle is the neutron star.}
\label{gamma_ion_fig}
\end{figure}

To conclude the discussion of active pulsars, Fig.\ \ref{gamma_ion_fig} shows the cell-averaged gamma factor for positrons in simulations A1 (left panel) and A2 (right panel). A large wedge of field lines at mid-latitudes above the return current layer do not support pair production in simulation A2, as already discussed.  Only electrons are present on these field lines, and thus there is a large black wedge in the right hand panel of Fig.\ \ref{gamma_ion_fig}, which contains only a few scattered positrons. One feature that simulations A1 and A2 have in common, though, is the acceleration of positrons in the current layer starting at the Y-line near the light-cylinder. This is evident in both panels of Fig.\ \ref{gamma_ion_fig} as the increase in gamma factor at the Y-line and in the current layer.

The electric field near the Y-line and in the current layer transforms electromagnetic energy to particle kinetic energy.  One unanswered question is what is the underlying reason for the presence of a strong accelerating field in the vicinity of the Y-line? It is not obvious that the electric field in pulsar PIC simulations is due to spontaneous magnetic reconnection, which is the focus of many idealized reconnection studies. Rather, it may be that the accelerating electric field in the vicinity of the Y-line is necessitated by the global structure of the magnetosphere. If this hypothesis is correct, then the electric field is required to ``straighten out" positron trajectories passing through the Y-line and make them more nearly radial. A better understanding of the accelerating electric field near the Y-line in PIC studies and how it depends on problem parameters is crucial for drawing correct comparisons between simulations and observations of high energy emission from pulsars.

\section{Particle Trajectories in the Current Layer and Near the Y-line}

\begin{figure}[!p]
\centering
\subfigure{\begin{overpic}
	[width=.9\textwidth]{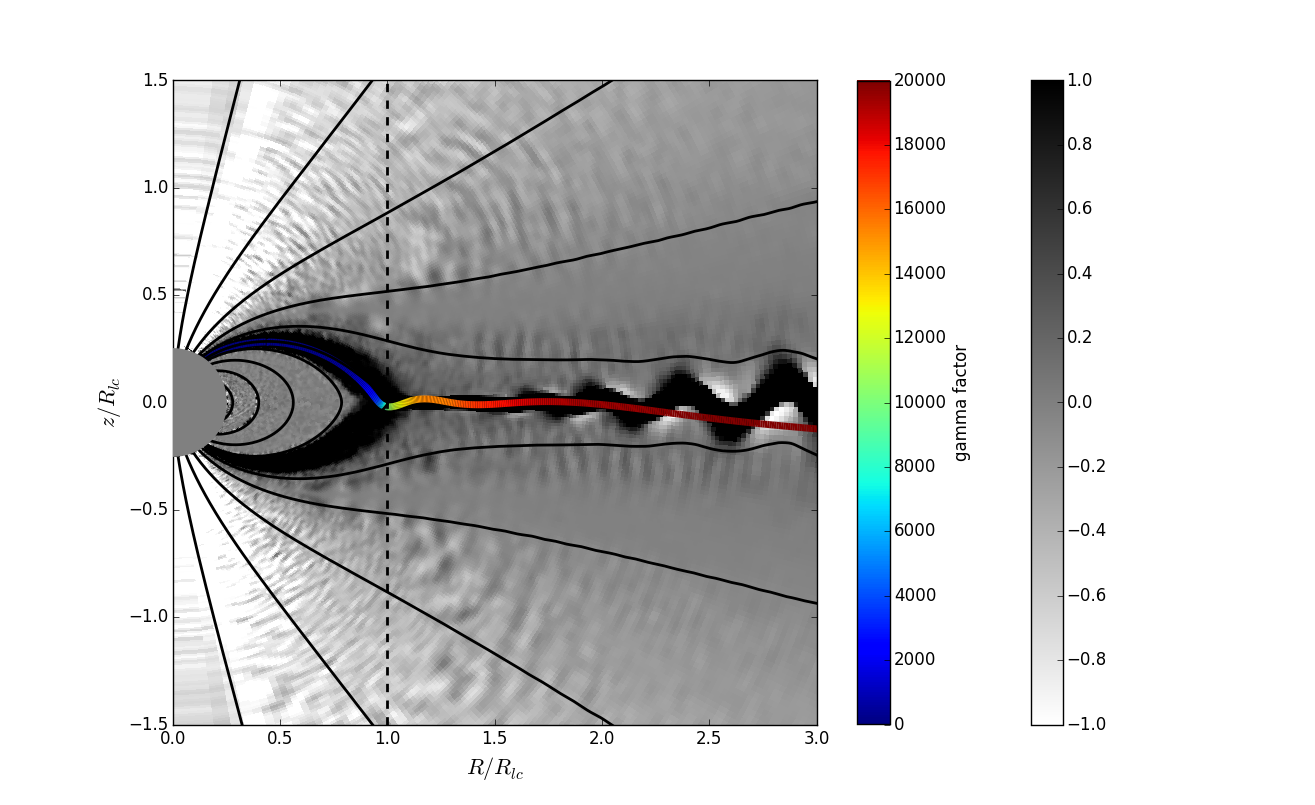}
	\put(48,50){\large \bf positron}
\end{overpic}}
\subfigure{\begin{overpic}
	[width=.9\textwidth]{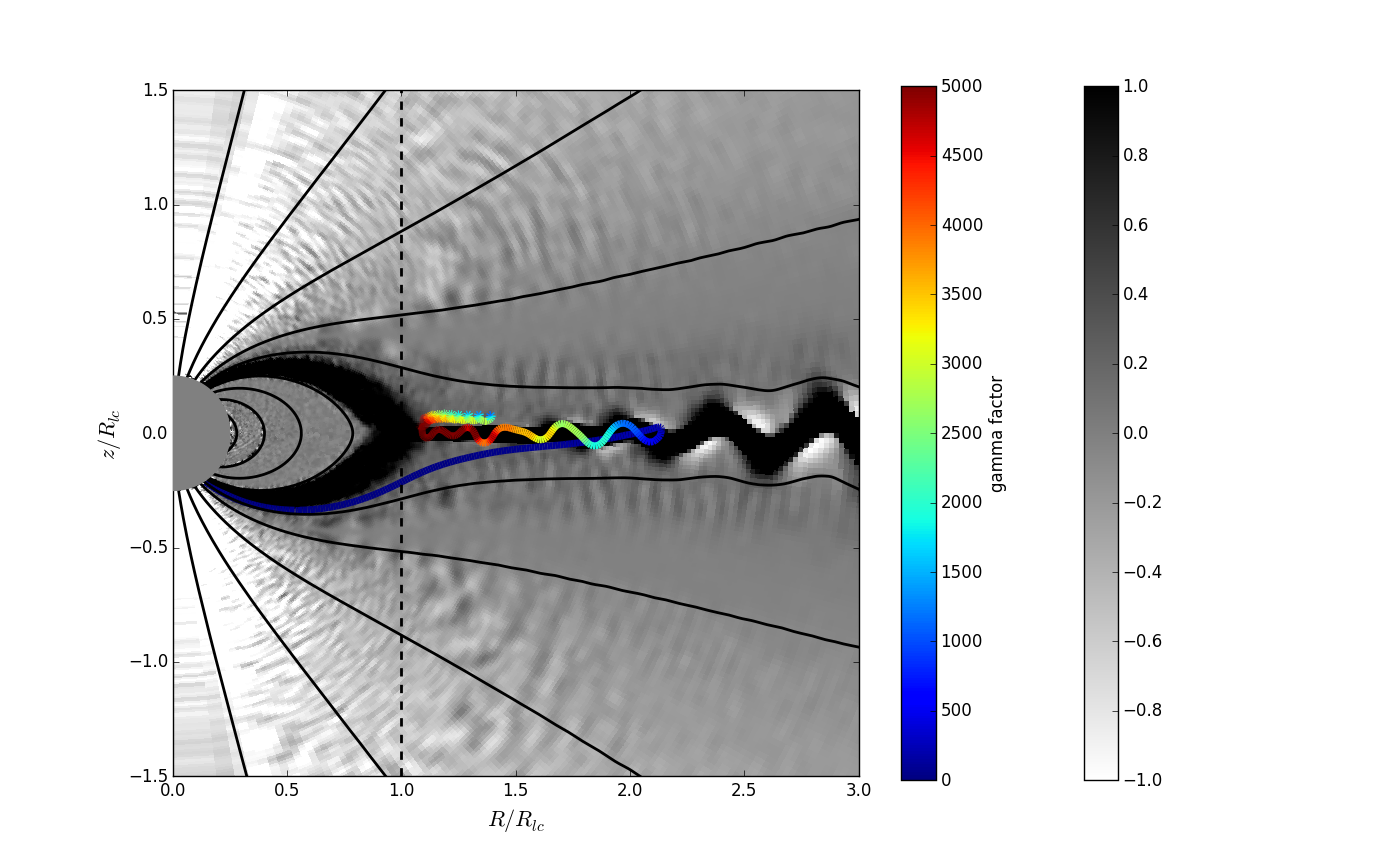}
	\put(48,50){\large \bf electron}
\end{overpic}}
\caption{Upper panel: Multicolored curve shows a sample positron trajectory that starts at the surface of the neutron star and passes through the Y-line into the current layer. Color indicates the gamma factor of the particle along its trajectory. The black and white background image shows the meridional current density normalized to a canonical value appropriate for the force-free solution. Lower panel: As above but for a sample electron trajectory that enters the current layer.}
\label{traj_fig}
\end{figure}

We now discuss particle trajectories on open field lines for an active pulsar, focusing particular attention on trajectories passing through the Y-line and current layer beyond the light cylinder. The electric field present at the Y-line and in the current layer acts as a particle filter, which accelerates positrons and electrons in opposite directions. Positrons entering the current layer are accelerated radially outward and undergo relativistic Speiser orbits \citep{UzdenskySpeiser}. If the current layer were exactly in the equatorial midplane, the radially-directed electric field would straighten positron trajectories making them more nearly radial and focusing positrons toward the midplane. However, the current layer itself is susceptible to kinking, which deflects particles away from the midplane. In contrast to positrons, electrons entering the current layer are accelerated inward towards the Y-line. These electrons cannot travel past the Y-line and are reflected at the corotation zone due to magnetic mirroring. The orbits of these electrons are complex, because they are radially trapped in the current layer near the Y-line. However, electrons can pop out of the current layer in the $\theta$ direction and become entrained in the wind above it. Such electrons undergo ExB drift in the nearly force-free electromagnetic fields just above the current layer.

The upper panel of Fig.\ \ref{traj_fig} shows the trajectory of a sample positron from simulation A1 that passes through the Y-line and into the current layer. As the positron approaches the Y-line, it is accelerated by the radially-directed electric field. It continues accelerating as it enters the current layer and initially starts to be focused into the midplane. However, for $R/R_{lc} \gtrsim 1.5$ the current layer is kinked, and the positron is deflected away from the midplane. Note that the kink pattern propagates radially at approximately the speed of light and is not stationary. Therefore, despite appearances, the positron whose trajectory spans many timesteps does not cross individual wiggles in Fig.\ \ref{traj_fig}. Rather, it is entrained inside the kinked current layer structure and propagates outward in phase with the kink. 

The lower panel of Fig.\ \ref{traj_fig} shows the trajectory of an electron that is absorbed into the current layer and accelerates inward toward the Y-line. Before reaching the Y-line, however, the electron pops out of the current layer in the $\theta$ direction, becomes entrained in the pulsar wind above it, and starts to undergo ExB drift in a predominantly radial direction. Note that the electron trajectory appears to end only because the end of the simulation has been reached. If the simulation were run for longer, the electron would continue to ExB drift outwards, either indefinitely or until it is reabsorbed into the current layer.

\begin{figure}[!t]
\centering
\includegraphics[width=.98\textwidth]{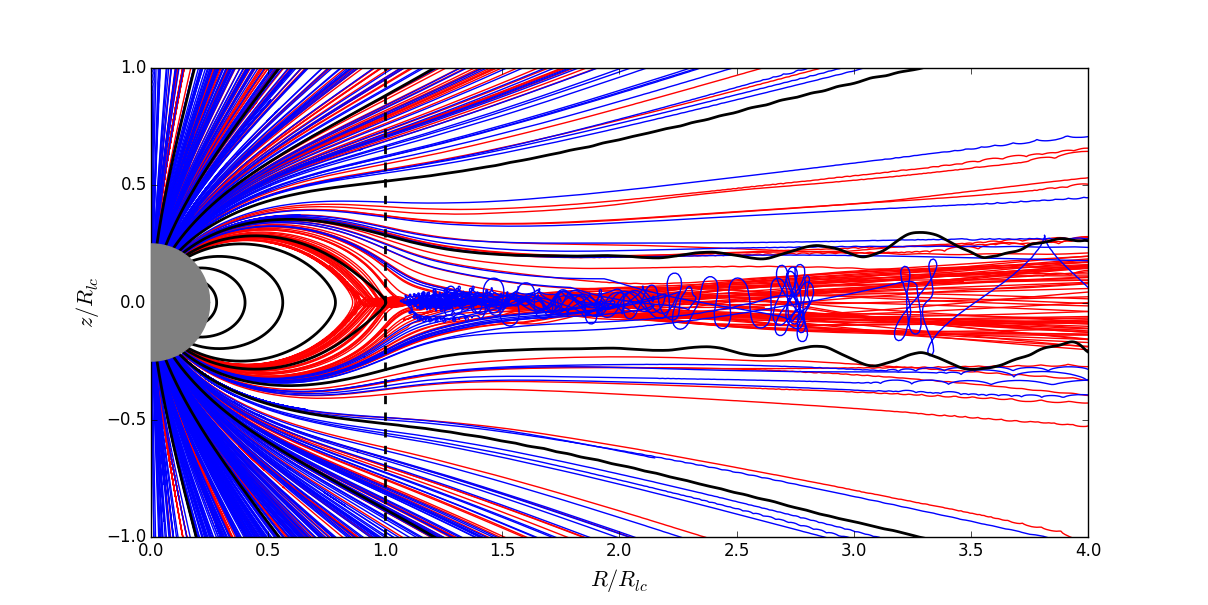}
\caption{Sample trajectories for electrons (blue) and positrons (red) that pass beyond $R = .75 R_\text{LC}$. Black curves are magnetic field lines and the gray sphere is the neutron star.}
\label{trajs_fig}
\end{figure}

Fig.\ \ref{trajs_fig} shows sample electron trajectories (blue) and positron trajectories (red) for simulations A1 that pass beyond $R = .75 R_\text{LC}$. Electrons and positrons that do not pass through the Y-line or enter the current layer follow trajectories along field lines in the meridional plane. The current layer is actually much narrower than it appears judging from the positron trajectories beyond $R = 1.5 R_\text{LC}$. This is because the current layer is kinked, and the latitudinal spread in positron trajectories traces the amplitude of the kink, not the width of the current layer. This is possible because the kink and the positrons in the current sheet propagate outward in phase at nearly the speed of light. 

Fig.\ \ref{trajs_fig} demonstrates filtering of particles by the electric field, which effectively removes all electrons from the current layer past a few light cylinder radii. We caution against reading too much into this, however, since pair multiplicities in global pulsar PIC simulations are unrealistically low for numerical reasons. Thus, electrons may not be entirely filtered out from the current layer in reality, and it would be interesting to increase the pair multiplicity in future simulations. In particular, it would be interesting to study how the pair multiplicity affects the strength of the electric field near the Y-line and in the current sheet.

\section{Performance Optimization}
{\it PICsar2D} has been optimized to take advantage of SIMD vectorization and cache reuse on modern processors. We focused on optimizing the mover, since this is the slowest step in the code. In fact, the mover is significantly slower than in a Cartesian PIC code, because of the additional coordinate transformations that are necessary in curvilinear coordinates. Fortunately, it is straightforward to achieve vectorization of the most time-intensive parts of the mover by adopting an Array of Structure of Arrays (AoSoA) data layout for the particles.

Storing the data in an AoSoA format is a standard technique used to achieve both good vectorization and cache localization for particle codes \citep{Decyk}. To understand why this is the case, it is useful to first consider the simpler example of an Array of Structures (AoS). In this case, the particle array using Fortran indexing notation (innermost index is the fastest) can be represented as a real array with dimensions p[PTL\_SIZE, N\_PTL, PTL\_TYPE]. Here, PTL\_SIZE is the size of a particle (e.g. if the particle only has 3 components of velocity and 3 of position then PTL\_SIZE = 6), N\_PTL is the maximum length of the particle list, and PTL\_TYPE determines how many particle species there are (e.g. PTL\_TYPE = 2 if there are only positrons and electrons). Because indices over the particle components change fastest, this leads to same memory layout as an array of particle structures. AoS leads to good cache reuse, since we typically want to update all of the components of a single particle before moving on to the next particle. However, it does not lead to good vectorization, because the same component (e.g.\ $x$-velocity) for particles that are adjacent in memory is at a stride of PTL\_SIZE.

To take advantage of SIMD vectorization and update multiple particles in one clock cycle, the individual components of position and velocity for different particles must have unit stride in memory. To achieve this, we can consider using a structure of arrays (SoA), which is the transpose of an AoS (individually for each species). Thus, the particle data is stored in memory as p[N\_PTL, PTL\_SIZE, PTL\_TYPE]. Vectorization is now possible, since individual components of position and velocity for different particles have unit stride in memory. However, good cache utilization is no longer possible if N\_PTL*PTL\_SIZE*SIZEOF(REAL) is greater than the size of cache in bytes, which is typically the case if N\_PTL is large.

To achieve both good vectorization and cache reuse, it is necessary to use an AoSoA memory layout for the particle data. In this case, the particle array can be represented in memory as p[VLEN,PTL\_SIZE, N\_PTL/VLEN, PTL\_TYPE]. Here, VLEN is a new free parameter that is related to the SIMD vector length of the machine. Notice that VLEN=1 corresponds to an SoA memory layout and VLEN=N\_PTL corresponds to an AoS memory layout. However, the real power of introducing a new parameter is due to the following reasons. First, if VLEN is a multiple of the machine vector length, then vectorization is possible for the same reason as in the AoS case. Second, good cache reuse only requires that VLEN*PTL\_SIZE*SIZEOF(REAL) is greater than the size of cache in bytes. Third, it is typically possible to choose VLEN to achieve both good vectorization and cache reuse in a machine-independent way (i.e.\ without having to change VLEN for running on a different processor type). In our experience, setting VLEN to have a size of 128 in units of bytes achieves the maximum possible vectorization without exceeding the cache size on current CPUs.

\begin{figure}[!p]
\includegraphics[width=.95\textwidth]{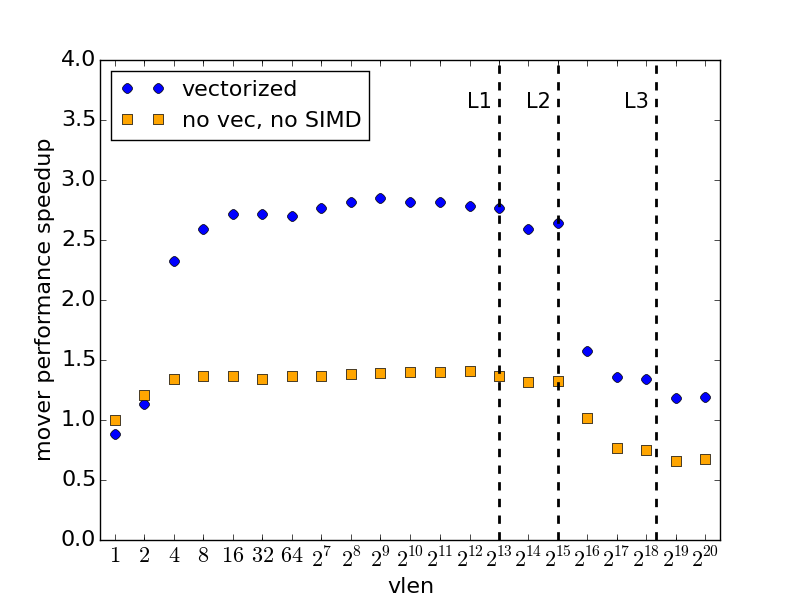}
\caption{Mover speedup (larger is better) on a single core as a function of the VLEN parameter using an AoSoA memory layout to store the particle data. The blue points show the case with SIMD vectorization turned on and the orange points with SIMD vectorization turned off (via compiler settings and compile-time options). The vertical dashed lines depict the L1, L2 and L3 cache sizes (the processor type is Sandy Bridge). Note the drop off in performance each time a cache boundary is exceeded. Performance is particularly bad when the vector length is so long that the particle data can no longer fit inside the L2 cache.}
\label{AoSoA_fig}
\end{figure}

Fig.\ \ref{AoSoA_fig} shows the relative mover performance when varying VLEN from $\text{VLEN}=1$ (AoS limit) to $\text{VLEN}=\text{N\_PTL}=2^{20}$ (SoA limit) in multiples of two for a pulsar test problem. At low values of VLEN, vectorization is inefficient, but at high values of VLEN, the cache size is exceeded, so memory accesses are inefficient. In between, $2^3 < \text{VLEN} < 2^{13}$, there is a plateau in performance when vectorization and cache reuse are both optimal.  Such a performance plateau will in general exist for each processor type, but the upper and lower extent of the plateau are determined by the SIMD vector length and cache size, which vary from processor to processor.

Storing the particles in memory using an AoSoA layout works well for vectorizing most of the work done by the mover without sacrificing cache performance. However, it does not vectorize the interpolation of field components to particle locations in the mover or the deposition of currents from individual particles to the grid in the current deposit step. In particular, the Villasenor-Buneman current deposition algorithm does not vectorize well, and the Esirkepov current deposition routine vectorizes only partially. Nevertheless, implementing an AoSoA layout for the particles in memory already improves code performance substantially.

\begin{figure}[!t]
\includegraphics[width=.95\textwidth]{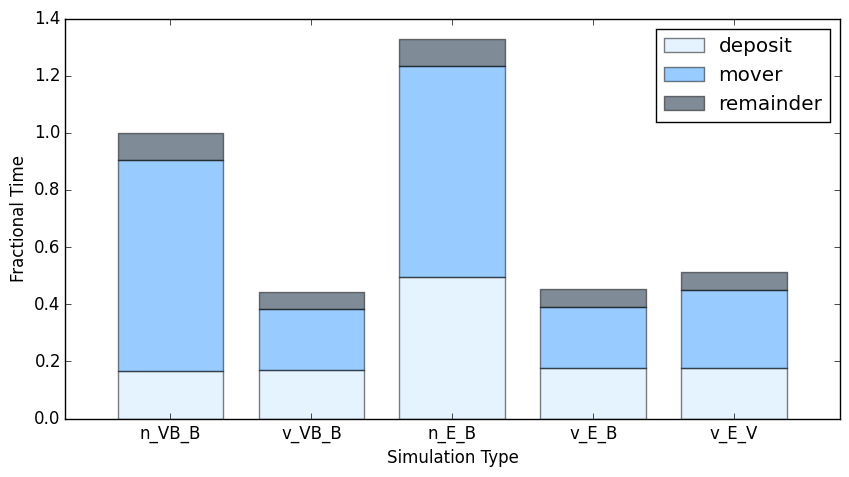}
\caption{Fractional time spent in different subroutines (smaller is better) for a pulsar test problem on a single core. The different segments of each bar depict the total time spent in the the current deposit (lower segment), mover (middle segment), and all other subroutines combined (upper segment). Different bars correspond to different types of simulations, and the identifying label under each bar is split into three parts. The first letter is ``v" if vectorization is turned on or ``n" if vectorization is turned off. The second string of letters determines the type of current deposition algorithm: ``E" for Esirkepov or ``VB" for Villasenor-Buneman. The last letter determines the type of mover: ``V" for Vay and ``B" for Boris. The processor type is Broadwell and $\text{VLEN} = 32$.}
\label{perf_hist}
\end{figure}

Fig.\ \ref{perf_hist} shows a histogram of the relative performance for various combinations of mover and deposit algorithms with and without SIMD vectorization using an AoSoA memory layout for the particles. SIMD vectorization dramatically reduces the amount of time spent by the code in the mover. In fact, the amount of time spent in the mover becomes about equal to the amount of time spent in the current deposit step, and the overall simulation time is reduced by over a factor of two.


\begin{thebibliography}{9}
\bibitem[Beloborodov(2008)]{Beloborodov} Beloborodov, A.~M.\ 2008, \apjl, 683, L41 
\bibitem[Belyaev(2015a)]{BelyaevPIC} Belyaev, M.~A.\ 2015, New Astronomy, 36, 37 
\bibitem[Belyaev(2015b)]{BelyaevPIC2} Belyaev, M.~A.\ 2015, \mnras, 449, 2759 
\bibitem[Belyaev \& Parfrey(2016)]{BelyaevParfrey} Belyaev, M.~A., \& Parfrey, K.\ 2016, \apj, 830, 119 
\bibitem[Boris(1970)]{Boris} Boris, J.~P. \ 1970, ``Relativistic plasma simulation-optimization of a hybrid code". Proceedings of the 4th Conference on Numerical Simulation of Plasmas. Naval Res. Lab., Washington, D.C. pp. 3 –- 67.
\bibitem[Cerutti et al.(2015)]{Cerutti} Cerutti, B., Philippov, A., Parfrey, K., \& Spitkovsky, A.\ 2015, \mnras, 448, 606 
\bibitem[Chen \& Beloborodov(2014)]{ChenBeloborodov} Chen, A.~Y., \& Beloborodov, A.~M.\ 2014, \apjl, 795, L22 
\bibitem[Contopoulos et al.(1999)]{CKF} Contopoulos, I., Kazanas, D., \& Fendt, C.\ 1999, \apj, 511, 351 
\bibitem[Decyk \& Singh(2014)]{Decyk} Decyk, V.~K., \& Singh, T.~V.\ 2014, Computer Physics Communications, 185, 708 
\bibitem[Esirkepov(2001)]{Esirkepov} Esirkepov, T.~Z.\ 2001, Computer Physics Communications, 135, 144 
\bibitem[Goldreich \& Julian(1969)]{GoldreichJulian} Goldreich, P., \& Julian, W.~H.\ 1969, \apj, 157, 869 
\bibitem[Gralla et al.(2016)]{GrallaPhilippov} Gralla, S.~E., Lupsasca, A., \& Philippov, A.\ 2016, \apj, 833, 258 
\bibitem[Krause-Polstorff \& Michel(1985)]{KrausePolstorffMichel} Krause-Polstorff, J., \& Michel, F.~C.\ 1985, \mnras, 213, 43P 
\bibitem[Michel \& Li(1999)]{MichelLi} Michel, F.~C., \& Li, H.\ 1999, \physrep, 318, 227 
\bibitem[Philippov et al.(2015a)]{Philippov3D} Philippov, A.~A., Spitkovsky, A., \& Cerutti, B.\ 2015, \apjl, 801, L19 
\bibitem[Philippov et al.(2015b)]{PhilippovGR} Philippov, A.~A., Cerutti, B., Tchekhovskoy, A., \& Spitkovsky, A.\ 2015, \apjl, 815, L19 
\bibitem[Spitkovsky \& Arons(2002)]{SpitkovskyPIC} Spitkovsky, A., \& Arons, J.\ 2002, Neutron Stars in Supernova Remnants, 271, 81 
\bibitem[Spitkovsky(2006)]{SpitkovskyFF} Spitkovsky, A.\ 2006, \apjl, 648, L51 
\bibitem[Sturrock(1971)]{Sturrock} Sturrock, P.~A.\ 1971, \apj, 164, 529 
\bibitem[Surmin et al.(2016)]{PICador} Surmin, I.~A., Bastrakov, S.~I., Efimenko, E.~S., et al.\ 2016, Computer Physics Communications, 202, 204 
\bibitem[Timokhin \& Arons(2013)]{TimokhinArons} Timokhin, A.~N., \& Arons, J.\ 2013, \mnras, 429, 20
\bibitem[Uzdensky et al.(2011)]{UzdenskySpeiser} Uzdensky, D.~A., Cerutti, B., \& Begelman, M.~C.\ 2011, \apjl, 737, L40
\bibitem[Vay(2008)]{Vay} Vay, J.-L.\ 2008, Physics of Plasmas, 15, 056701 
\bibitem[Villasenor \& Buneman(1992)]{VillasenorBuneman} Villasenor, J., \& Buneman, O.\ 1992, Computer Physics Communications, 69, 306
\bibitem[Yee(1966)]{Yee} Yee, K.\ 1966, IEEE Transactions on Antennas and Propagation, 14, 302  
\end{thebibliography}
\end{document}